\documentclass[10pt]{cernrep_a}
\usepackage{graphicx}
\usepackage{floatflt}

\newcommand{\Pt}{{P_t}}
\newcommand{\dphi}{\Delta\phi}
\newcommand{\phigj}{\phi_{(\gamma,jet)}}

\newcommand{\la}{\langle}
\newcommand{\ra}{\rangle}
\newcommand{\gpj}{~"$\gamma+Jet$"~}
\newcommand{\rrr}{\to} 

\newcommand{\Db}{$\Pt(O\!+\!\eta>5)$}
\newcommand{\ptgj}{$~\Pt^{\gamma}$ and $\Pt^{Jet}~$}
\newcommand{\ptg}{$\Pt^{\gamma}$}
\newcommand{\Ptg}{\Pt^{\gamma}}

\newcommand{\Fptgj}{(\Pt^{\gamma}\!-\!\Pt^{jet})/\Pt^{\gamma}}

\newcommand{\aaa}{\hspace*{.39cm}}

\suppressfloats[!]

\begin{document}
\thispagestyle{empty}

\vskip-5mm

\begin{center}
{\Large JOINT INSTITUTE FOR NUCLEAR RECEARCH}
\end{center}

\vskip10mm

\begin{flushright}
JINR Preprint \\
E2-2000-252 \\
hep-ex/0011013
\end{flushright}

\vspace*{3cm}

\begin{center}
\noindent
{\Large{\bfseries Jet energy scale setting with \gpj events at LHC energies. \\[5pt]
Event rates, $\Pt$ structure of jet.}}\\[5mm]
{\large D.V.~Bandourin$^{1\,\dag}$, V.F.~Konoplyanikov$^{2\,\ast}$, 
N.B.~Skachkov$^{3\,\dag}$}

\vskip 0mm

{\small
{\it
E-mail: (1) dmv@cv.jinr.ru, (2) kon@cv.jinr.ru, (3) skachkov@cv.jinr.ru}}\\[3mm]
$\dag$ \large \it Laboratory of Nuclear Problems \\
\hspace*{-4mm} $\ast$ \large \it Laboratory of Particle Physics
\end{center}

\vskip 9mm
\begin{center}
\begin{minipage}{150mm}
\centerline{\bf Abstract}
\noindent
In this paper the study of \gpj events is continued,
aimed at jet energy scale setting and hadron calorimeter calibration
at LHC energies. The event number distribution over \ptg and $\eta^\gamma$
is presented. The features of \gpj events in the barrel region of
CMS detector ($|\eta|<1.4$) are exposed. $\Pt$ structure of the region
in the $\eta-\phi$ space inside and beyond jet is also shown.
\end{minipage}
\end{center}

\newpage

\setcounter{page}{1}
\section{INTRODUCTION} 
In this paper we continue the study of \gpj process, started in [1]
aimed at jet energy scale setting and hadron calorimeter calibration
at LHC energies.

This article is organized in the following way. In Section 2
we shall estimate the event number distribution dependence on a value of
"back-to-back" $\phigj$ angle between the direct photon $\vec{\Pt}^{\gamma}$
and jet $\vec{\Pt}^{Jet}$ and
on the $\Pt$ of the initial state radiation (ISR), i.e. $\Pt^{ISR}$ value.
$\Pt^{\gamma}$ and $\eta^{\gamma}$ dependence of rates is also
studied in this section.
Estimation of \gpj event rates for Barrel (HB), Endcap (HE) and Forward
(HF) parts of hadron calorimeter (HCAL) is performed in the last subsection
 of Section 2. In Section 3 we consider the \gpj event features in HB region.
In particular, we study the $\Pt^{clust}_{CUT}~$ parameter influence on
the selection of events with a good photon and jet $\Pt$ balance and 
also the problem of selection of events with suppressed ISR activity. The next subsections
are devoted to the $\Pt$ structure of jets in $\eta-\phi$ space
and to the particles $\Pt$ distribution inside and beyond  the jet.

\section{EVENT RATES FOR DIFFERENT $\Pt^{\gamma}$ AND $\eta^{\gamma}$
 INTERVALS.}

\subsection{Event number distribution dependence  on "back-to-back"
$\phigj$ angle and on $\Pt^{ISR}$ values. }                           

The definitions of observable physical variables introduced in
Section 3.2 of [1] allow us to make the first step in the study
 of a possible way
to select the events with a good $\Pt^{\gamma}$ and $\Pt^{Jet}$ balance.
Here we shall study the spectrum of $\Pt{56}$ variable
(that is proportional to $\Pt^{ISR}$ up to the value of intrinsic parton
 transverse momentum inside a proton)
in signal events with direct photons and this spectrum
dependence on some of experimentally
measurable variables. For this reason the results of $ 10^7$ \gpj events
generation for
every of four $\Pt^{\gamma}$ intervals (40--50, 100--120, 200--240, 300--360
$GeV/c$) by use of PYTHIA with account of
2 QCD subprocesses $~q\bar{q}\rrr g+\gamma~$
 and $~q\,g\rrr q+\gamma~$ were analyzed. For this analysis the observables
 and the ``Selection 1'' cuts (1--7), defined in Sections 3.2 of [1],
with the following cut parameters:
\begin{eqnarray}
\Pt^{isol}_{CUT}=5\;GeV/c; \quad
{\epsilon}^{\gamma}_{CUT}=7\%; \quad
\dphi<15^{\circ}; \quad
\Pt^{clust}_{CUT}=20\;GeV/c.
\end{eqnarray}
were used.
The value of the luminosity was taken to be $L=10^{33}\,cm^2sec^{-1}$
(low LHC luminosity).

In Tables 1, 2 and 4, 5 we present $\Pt56$ spectra for
two most illustrative cases of
$\Pt^{\gamma}$ intervals: $40<\Pt^{\gamma}<50 ~GeV/c$ (Table 1 and 2) and
$200<\Pt^{\gamma}<240 ~GeV/c$ (Table 4 and 5). The distributions of the 
number of events for integrated luminosity $L_{int}=3\,fb^{-1}$
in different intervals of $\Pt56$ ($\la k_T\ra$ was taken to be
fixed at PYTHIA default value, i.e. $\la k_T\ra=0.44\,GeV/c$)
 and for different ``back-to-back'' angle intervals
$\phigj=180^\circ \pm \dphi~$ ($\dphi=
15^\circ,\,10^\circ$ and $5^\circ$
as well as without any restriction on
$\dphi$, i.e. for the whole $\phi$ interval
$\Delta\phi=180^\circ$)
\footnote{The value $\Delta\phi=5^\circ$
coincides with one CMS  HCAL tower size in the $\phi$-plane.
and, thus, defines the precision of experimental registration 
of the jet $\vec{\Pt}$ vector.}
 are given there.
Tables 1 and 2
correspond to the case $\Pt^{clust}<20\,GeV/c$
and serve as an illustration since it is rather a 
weak cut condition, while
Tables~ 4 and 5 correspond to a more
restrictive selection cut value $\Pt^{clust}_{CUT}=5\,GeV/c$.

Firstly from the last summary lines of Tables 1--4
we can make a general conclusion about the $\dphi$-dependence
of the events spectrum.
Thus, in the case of weak restriction  $\Pt^{clust}\!<20 ~GeV/c$
we can see from Table 1 that for the interval of $40\leq \Pt^{\gamma}\leq 50\, GeV/c$
there are about 66$\%$ of events concentrated
in the $\Delta\phi<15^\circ$ range, while 32$\%$ of events are
in the $\Delta\phi<5^\circ$ range.
At the same time the analogous summary line of Table 2
shows us that at higher values
$200\leq \Pt^{\gamma}\leq 240\, GeV/c$ the events spectrum
moves to the small $\dphi$ region: more than 99$\%$ of events have
$\Delta\phi<15^\circ$ and 79$\%$ of them have $\Delta\phi<5^\circ$.
%
\def\baselinestretch{0.98}
\begin{table}[htbp]
\begin{center}
\vskip-1.2cm
\caption{Number of events dependence on $\Pt56$ and
$\Delta\phi$ for}
\vskip-3pt
{\footnotesize $40\leq Pt^{\gamma}\leq 50 \, GeV/c$ and
$\Pt^{clust}_{CUT}= 20 \, GeV/c$ for $L_{int}$=3$fb^{-1}$}
\vskip.2cm
\begin{tabular}{||c||r|r|r|r||} \hline \hline
 $\Pt{56}$&\multicolumn{4}{|c||}{ $\dphi_{max}$} \\\cline{2-5}
 $(GeV/c)$  &$180^\circ$&$15^\circ$ & $10^\circ$&$5^\circ$ \\\hline\hline
    0 --   5 &   1103772 &   1049690 &   1006627 &    849706 \\\hline
    5 --  10 &   1646004 &   1564393 &   1403529 &    812304 \\\hline
   10 --  15 &   1331589 &   1122473 &    771060 &    380122 \\\hline
   15 --  20 &    992374 &    568279 &    365329 &    179767 \\\hline
   20 --  25 &    725537 &    282135 &    183406 &     91113 \\\hline
   25 --  30 &    559350 &    169186 &    112308 &     58395 \\\hline
   30 --  40 &    911942 &    265961 &    178048 &     89867 \\\hline
   40 --  50 &    388950 &     94112 &     62068 &     31000 \\\hline
   50 -- 100 &     91248 &     19442 &     12973 &      6234 \\\hline
  100 -- 300 &        34 &         0 &         0 &         0 \\\hline
  300 -- 500 &         0 &         0 &         0 &         0 \\\hline \hline
    0 -- 500 &   7750799 &   5135671 &   4095348 &   2498507 \\\hline \hline
\end{tabular}
\vskip0.2cm
\caption{Number of events dependence on $\Pt56$ and
$\Delta\phi$ for}
\vskip-3pt
{\footnotesize $200\leq Pt^{\gamma}\leq 240 \, GeV/c$ and
$\Pt^{clust}_{CUT}= 20 \, GeV/c$ for $L_{int}$=3$fb^{-1}$}
\vskip0.2cm
\begin{tabular}{||c||r|r|r|r||} \hline \hline
 $\Pt{56}$  &\multicolumn{4}{c||}{ $\dphi_{max}$} \\\cline{2-5}
 $(GeV/c)$  &\aaa $180^\circ$\aaa&\aaa$15^\circ$\aaa&\aaa$10^\circ$\aaa&\aaa$5^\circ$\aaa \\\hline\hline
    0 --   5 &      1429 &      1429 &      1427 &      1380 \\\hline
    5 --  10 &      3266 &      3266 &      3264 &      3150 \\\hline
   10 --  15 &      3205 &      3205 &      3200 &      3069 \\\hline
   15 --  20 &      2827 &      2827 &      2819 &      2618 \\\hline
   20 --  25 &      2409 &      2408 &      2393 &      1918 \\\hline
   25 --  30 &      2006 &      2006 &      1982 &      1300 \\\hline
   30 --  40 &      2608 &      2605 &      2533 &      1411 \\\hline
   40 --  50 &      1237 &      1230 &      1067 &       586 \\\hline
   50 -- 100 &      1066 &      1018 &       842 &       536 \\\hline
  100 -- 300 &       313 &       307 &       293 &       221 \\\hline
  300 -- 500 &         0 &         0 &         0 &         0 \\\hline \hline
    0 -- 500 &     20366 &     20301 &     19820 &     16189 \\\hline \hline
\end{tabular}
\vskip0.2cm
\caption{Number of events dependence on $\dphi_{max}$ and on
$\Pt^{\gamma}$ for $L_{int}=3\,fb^{-1}$.}
\vskip-3pt
{\footnotesize $\Pt^{clust}_{CUT}=20 ~GeV/c$ (summary)}
\vskip0.2cm
\begin{tabular}{||c||r|r|r|r||} \hline \hline
 $\Pt^{\gamma}$  &\multicolumn{4}{c||}{ $\dphi_{max}$} \\\cline{2-5}
 $(GeV/c)$  &$180^\circ$&$15^\circ$&$10^\circ$&$5^\circ$ \\\hline\hline
 40 -- 50  &7750799 &   5135671 &   4095348 &   2498507 \\\hline
100 -- 120 & 323766 &    297323 &    258691 &    176308 \\\hline
200 -- 240 &  20366 &     20301 &     19820 &     16189 \\\hline
300 -- 360 &   3638 &      3638 &      3627 &      3323 \\\hline\hline
\end{tabular}
\end{center}
\end{table}
\begin{table}[htbp]
\begin{center}
\caption{Number of events dependence on $\Pt56$ and
$\Delta\phi$ for}
\vskip-3pt
{\footnotesize $40\leq Pt^{\gamma}\leq 50 \, GeV/c$ and
$\Pt^{clust}_{CUT}= 5 \, GeV/c$ for $L_{int}$=3$fb^{-1}$}
\vskip0.2cm
\begin{tabular}{||c||r|r|r|r||} \hline \hline
 $\Pt{56}$  &\multicolumn{4}{c||}{ $\dphi_{max}$} \\\cline{2-5}
 $(GeV/c)$  &$180^\circ$&$15^\circ$&$10^\circ$&$5^\circ$ \\\hline\hline
    0 --   5 &    331522 &    331321 &    329876 &    295759 \\\hline
    5 --  10 &    319153 &    318581 &    299960 &    187089 \\\hline
   10 --  15 &     88603 &     82586 &     60537 &     32335 \\\hline
   15 --  20 &     21244 &     15327 &     11663 &      6924 \\\hline
   20 --  25 &      8101 &      5681 &      4639 &      2992 \\\hline
   25 --  30 &      4739 &      3395 &      2823 &      1949 \\\hline
   30 --  40 &      3495 &      2790 &      2555 &      1714 \\\hline
   40 --  50 &      1647 &      1277 &      1042 &       471 \\\hline
   50 -- 100 &       101 &        67 &        67 &        67 \\\hline
  100 -- 500 &         0 &         0 &         0 &         0 \\\hline \hline
    0 -- 500 &    778606 &    761026 &    713161 &    529299 \\\hline \hline
\end{tabular}
\vskip0.2cm
\caption{Number of events dependence on $\Pt56$ and
$\Delta\phi$ for}
\vskip-3pt
{\footnotesize $200\leq Pt^{\gamma}\leq 240 \, GeV/c$ and
$\Pt^{clust}_{CUT}= 5 \, GeV/c$ for $L_{int}$=3$fb^{-1}$}
\vskip0.2cm
\begin{tabular}{||c||r|r|r|r||} \hline \hline
 $\Pt{56}$  &\multicolumn{4}{c||}{ $\dphi_{max}$} \\\cline{2-5}
 $(GeV/c)$  &\aaa $180^\circ$\aaa&\aaa$15^\circ$\aaa&\aaa$10^\circ$\aaa&\aaa$5^\circ$\aaa \\\hline\hline
    0 -   5 &       369 &       369 &       369 &       369 \\\hline
    5 -  10 &       563 &       563 &       563 &       562 \\\hline
   10 -  15 &       217 &       217 &       217 &       217 \\\hline
   15 -  20 &        56 &        56 &        56 &        56 \\\hline
   20 -  25 &        20 &        20 &        20 &        18 \\\hline
   25 -  30 &         9 &         9 &         9 &         7 \\\hline
   30 -  40 &         7 &         7 &         7 &         6 \\\hline
   40 -  50 &         6 &         6 &         6 &         5 \\\hline
   50 - 100 &        10 &        10 &        10 &        10 \\\hline
  100 - 300 &         8 &         8 &         8 &         8 \\\hline
  300 - 500 &         0 &         0 &         0 &         0 \\\hline \hline
    0 - 500 &      1264 &      1264 &      1264 &      1257 \\\hline \hline
\end{tabular}
\vskip0.2cm
\caption{Number of events dependence on $\dphi_{max}$ and on
 $\Pt^{\gamma}$ for $L_{int}=3\,fb^{-1}$.}
\vskip-3pt
{\footnotesize $\Pt^{clust}_{CUT}= 5 \, GeV/c$ (summary)}
\vskip0.2cm
\begin{tabular}{||c||r|r|r|r||} \hline \hline
 $\Pt^{\gamma}$  &\multicolumn{4}{c||}{ $\dphi_{max}$} \\\cline{2-5}
 $(GeV/c)$  &$180^\circ$&$15^\circ$&$10^\circ$&$5^\circ$ \\\hline\hline
\hline
 40 -- 50  &778606 &    761026 &    713161 &    529299 \\\hline
100 -- 120 & 22170 &     22143 &     22038 &     20786 \\\hline
200 -- 240 &  1264 &      1264 &      1264 &      1257 \\\hline
300 -- 360 &   212 &       212 &       212 &       212 \\\hline
\end{tabular}
\end{center}
\end{table}

\def\baselinestretch{1.0}

A tendency of very rapid concentration of the signal \gpj event
distributions with the growth of $\Pt^{\gamma}$ in a rather narrow
back-to-back angle interval $\Delta\phi<15^\circ$
becomes more distinctive with a more restrictive cut
$\Pt^{clust}_{CUT}= 5\,GeV/c$ (see Tables 4 and 5). From
the last summary line of the Tables 3 we see that in the case of
$40\leq \Pt^{\gamma}\leq 50\, GeV/c~$ more than
$97\%$ of the events have $\Delta\phi<15^\circ$, while $~68\%$ of them are
in the $\Delta\phi<5^\circ$ range.
At $200\leq \Pt^{\gamma}\leq 240\, GeV/c$
(see Table 5) more than  $99\%$ of the events that undergo this cut
have $\Delta\phi<5^\circ$.
It means that imposing $\Pt^{clust}_{CUT}= 5 ~GeV/c$ and suppressing 
cluster or mini-jet activity, we choose the events with a clean
``back-to-back'' (within 15$^\circ$) topology of $\gamma$ and jet orientation.

So, one can conclude that PYTHIA simulation predicts that
at LHC  energies most of the \gpj events after
imposing $\Pt^{clust}_{CUT}= 20 ~GeV/c$ may have the vectors
$\vec{\Pt}^{\gamma}$ and $\vec{\Pt}^{jet}$ being back-to-back
within $\Delta\phi<15^\circ$.  The cut $\Pt^{clust}_{CUT}= 5 \; GeV/c$
improves this tendency
\footnote{The enlarging of \ptg values produces the same effect as it is seen 
from Tables 2 and 4 and
will be demonstrated in more detail in our following papers [2, 3].}.

It is worth mentioning that this picture
reflects the predictions of one of the generators which
are based on the approximate  LO values for
the cross section. It may change if the
next-to-leading order or soft physics 
\footnote{We thank E.~Pilon and J.Ph.~Jouliet for the information
about new FNAL data on this subject and on the importance of NLO corrections
and soft physics effects.}
effects are included
and also after more real experimental data would be collected for 
comparison with a theory.

The other lines of the same Tables 1, 2 and 4, 5 contain the information
about the $\Pt56$ (see formula (3) of [1]) spectrum
that in reality corresponds approximately, up to the fixed value of
$\langle k_T \rangle=0.44\;GeV/c$, 
to the $\Pt^{ISR}$ spectrum.

From the comparison of Table 1 with  Table 4 one can
conclude that the most populated part of $\Pt56$ (or $\Pt^{ISR}$)
spectrum reduces practically by twice
with restricting $\Pt^{clust}_{CUT}$. So for $\dphi_{max}=15^\circ$ we
see that it drops
from $0<\Pt{56}<50\;GeV/c$ $\!$ for $\!$ $\Pt^{clust}_{CUT}=20\;GeV/c$ to a
more $\!$ narrow interval $\,$ of $\!$
 $0<\Pt{56}<20\,GeV/c$ $\!$ for $\!$ the $\Pt^{clust}_{CUT}=5\,GeV/c$ case.
At higher $\Ptg$ interval for the same value $\dphi_{max}=15^\circ$
about the same factor 2 of $\Pt56$ spectrum  reduction
(from an interval $0<\Pt56<50~ GeV/c$ for $\Pt^{clust}_{CUT}=20\,GeV/c$
to a $0<\Pt56<20 ~GeV/c$ for $\Pt^{clust}_{CUT}=5 ~GeV/c$)
can be found from the comparison of Table 2 with Table 5.

From the same tables one can also see that $\Pt56$ spectrum becomes
slightly wider for higher values of
$\Pt^{\gamma}$: the dominant part of this spectrum
 in the interval $40\leq \Pt^{\gamma}\leq 50 ~ GeV/c$
 spreads in the case of $\Pt^{clust}_{CUT}=5 ~GeV/c$ cut
mainly within the interval $0<\Pt56<10 ~GeV/c$ with the maximum peak at
$0<\Pt{56}<5 ~GeV/c$.
At higher $\Pt^{\gamma}$ interval
$200\leq \Pt^{\gamma}\leq 240~ GeV/c$, $\Pt56$ spectrum spreads out
in a  wider interval of $0<\Pt{56}<15 ~GeV/c$ having the peak of its
 maximum at $5<\Pt{56}<10 ~GeV/c$.

Thus, we can summarize that the PYTHIA  generator used here predicts
the values of $\Pt^{ISR}$ spectrum
to increase while growing $\Pt^{\gamma}$, but its contribution
can be reduced by imposing a restrictive cut on the value
of $\Pt^{clust}$ (for more details see Section 3 and the following papers [2, 3]).

Since the last lines in Tables 1, 2 and 4, 5 contain an important
information on $\Delta\phi$ dependence of
the total number of events, we supply these tables with the
summarizing
Tables 3 and 6. They include more intervals of
$\Pt^{\gamma}$ and contain analogous numbers of events that can be collected
in different $\Delta\phi$ intervals for two different
 $\Pt^{clust}$ cuts at $L_{int}=3\,fb^{-1}$ (one month
of LHC continuous running at low luminosity).

\subsection{$\Pt^{\gamma}$ and $\eta^{\gamma}$ dependence of rates.}
\def\baselinestretch{1.0}
\begin{flushleft}
\parbox[r]{.6\linewidth}{
Since there are two main objects for  experimental registration
during the calibration procedure, namely a photon and a jet, we shall present
here the number of events predicted by PYTHIA simulation with cuts defined
by (1) for different
intervals of $\Pt^{\gamma}$ and $\eta^{\gamma}$.
The lines of Table 7 correspond to $\Pt^{\gamma}$-intervals while
the columns --- to $\eta^{\gamma}$-intervals .
The last column of this table contains the total number
of events in the whole ECAL
(at $L_{int}=3\,fb^{-1}$) $\eta^{\gamma}$-region
$0<|\eta^{\gamma}|<2.61$ for a given $\Pt^{\gamma}$-interval.
We see from  here that the events number decreases fastly
with the $\Pt^{\gamma}$ growth
(by more than 50$\%$ for each following interval) while for the fixed
$\Pt^{\gamma}$-
interval there are no big changes with the variation of $\eta^{\gamma}$.
Since ~$L_{int}=3\,fb^{-1}$ corresponds approximately to
one month of LHC continuous running, we conclude that these
rates demonstrate that, in principle, there would 
}
\end{flushleft}
\begin{flushright}
\begin{figure}[htbp]
\vskip-7.9cm
  \hspace{8.1cm} \includegraphics[width=6.5cm,height=5.5cm]{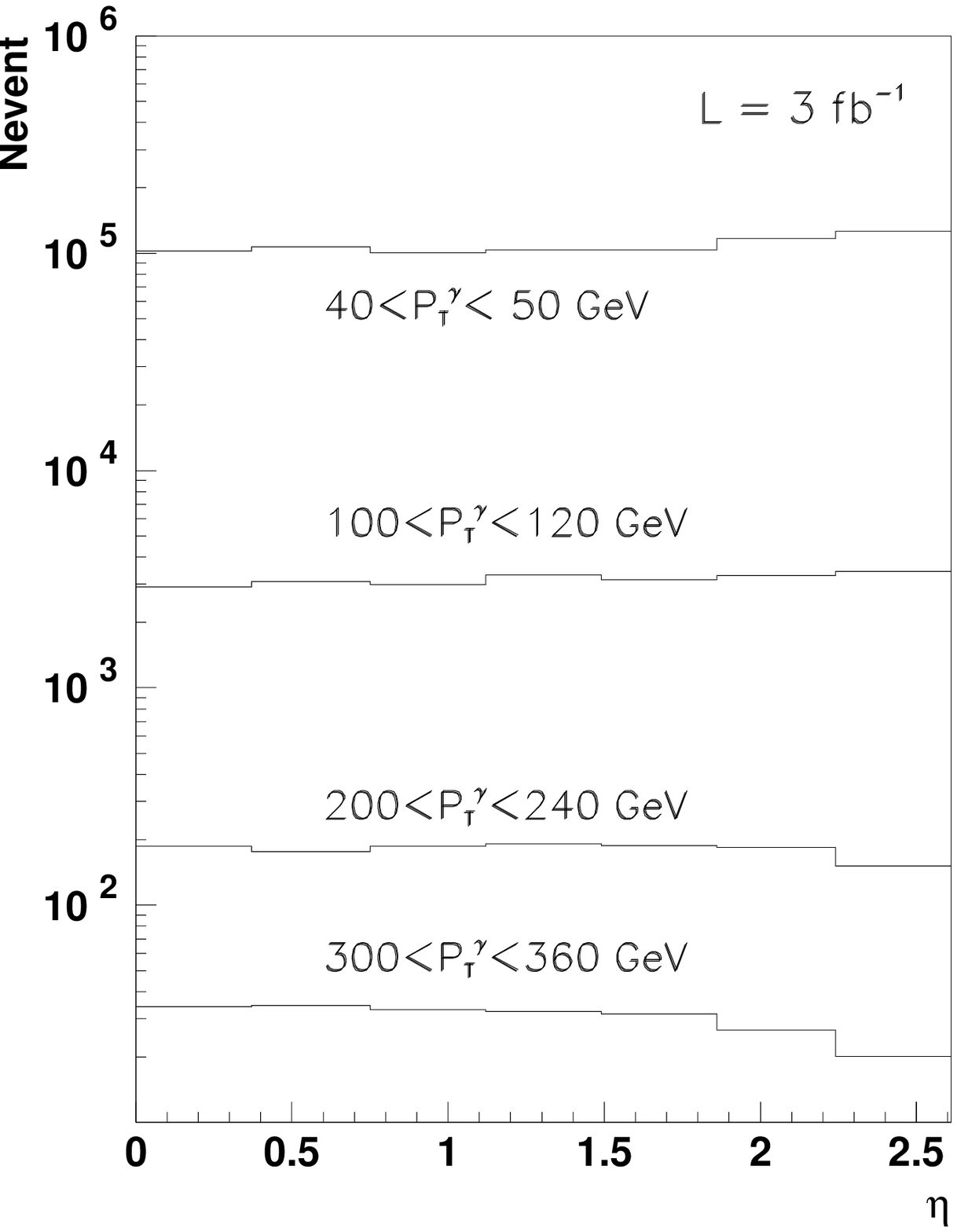}
    \label{fig6}
\end{figure}
\end{flushright}
\vspace{-1.4cm}
\hspace*{8.5cm} {\footnotesize {Fig.~1: $\eta$-dependence of rates for\\}}
\hspace*{9.4cm} {\footnotesize {different $\Pt^{\gamma}$-intervals}}
~\vspace*{0.3cm}
\noindent
\def\baselinestretch{0.99}
\begin{table}[h]
\begin{center}
\caption{Rates for $L_{int}=3\,fb^{-1}$ for different intervals of $\Pt^{\gamma}$ and $\eta^{\gamma}$
($\Pt^{clust}_{CUT}= 5 \, GeV/c$ and $\Delta\phi \leq 15^\circ$).}
\vskip-0.2cm
\begin{tabular}{||c||r|r|r|r|r|r|r||r||} \hline \hline
$\Pt^{\gamma}$ &\multicolumn{7}{c||}{$\eta^{\gamma}$~~ intervals}
&all ~~$\eta^{\gamma}$ \\\cline{2-9}
$(GeV/c)$ & 0.0-0.4 & 0.4-0.7 & 0.7-1.1 & 1.1-1.5 & 1.5-1.9 & 1.9-2.2 & 2.2-2.6 &0.0-2.6
 \\\hline
 \hline
 40 --  50& 102656& 107148& 100668& 103903& 103499& 116674& 126546& 761027\\\hline
 50 --  60&  43905&  41729&  41074&  45085&  42974&  47640&  50310& 312697\\\hline
 60 --  70& 18153 & 18326 & 19190&  20435&  20816 & 19432&  23650& 140005\\\hline
 70 --  80&   9848 & 10211&  9963 & 10166&  9951 & 11397 & 10447&  71984\\\hline
 80 --  90&   5287&  5921&  5104&  5823&  5385&  6067&  5923&  39509\\\hline
 90 -- 100&   2899&  3033&  3033&  3326&  3119&  3265&  3558&  22234\\\hline
100 -- 120&   2908&  3091&  2995&  3305&  3133&  3282&  3429&  22143\\\hline
120 -- 140&   1336  & 1359  & 1189  & 1346  & 1326  & 1499  & 1471&  9525\\\hline
140 -- 160&    624&   643&   626&   674&   706&   614&   668&  4555\\\hline
160 -- 200&    561&   469&   557&   555&   519&   555&   557&  3774\\\hline
200 -- 240&   187   & 176   & 186   & 192   & 187   & 185   & 151  & 1264\\\hline
240 -- 300&   103&    98&    98&    98   & 100&    92&    74&   665\\\hline
300 -- 360&     34&    34&    33&    32&    31&    27&    20&   212\\\hline \hline
 40 -- 360& 188517& 192274& 184734& 194957& 191761&  210742&  226819& 1389484\\\hline
\hline
\end{tabular}
\end{center}
\end{table}
\vskip-0.5cm
\noindent
 be quite enough number of events for the jet mass scale
setting and  successful calibration of HCAL.
Fig.~1 gives a compact illustration of rates for four $\Pt^{\gamma}$-intervals,
considered here, 
\def\baselinestretch{1.0}

\begin{table}[htpb]
and it shows a very weak dependence on $\eta^{\gamma}$
(or better to speak about their independence on $\eta^{\gamma}$).
\subsection{Estimation of \gpj events rates for HB, HE and HF regions.} 

Since the jet is a wide spread object, the $\eta^{jet}$ dependence of rates
for different $\Pt^{\gamma}$-intervals will be presented in a different way
than in Table 7. Namely, Tables 8--11 include the rates of events for different
$\eta^{jet}$ intervals, covered by the barrel, endcap and forward
(HB, HE and HF) parts of HCAL and for different
$\Pt^{Jet}$ intervals (for a case of $L_{int}=3\,fb^{-1}$).
The selection cuts are taken as those of Section 3.2 of [1] specified
by the following values of cuts parameters:
\\[-15pt]
\begin{center}
\caption{Selection 1. $\Delta \Pt^{jet} / \Pt^{jet} = 0.00$}
\vskip0.2cm
\begin{tabular}{||c||c|c|c|c|c||} \hline \hline
$\Pt^{\gamma}$&   HB   &   HB+HE &     HE  &   HE+HF &     HF\\\hline \hline
 40 --  50& 260259&  211356&  141759&  102299& 45354 \\\hline
 50 --  60& 108827& 89126& 55975& 41553& 17216 \\\hline
 60 --  70&  49585& 40076& 25172& 18153&  7019 \\\hline
 70 --  80&  25506& 20897& 12881&  9679&  3021 \\\hline
 80 --  90&  14083& 11720&  7529&  4873&  1304 \\\hline
 90 -- 100&   7261&  7054&  4142&  2924&   853 \\\hline
100 -- 120&   7703&  6913&  4013&  2926&   588 \\\hline
120 -- 140&   3372&  2977&  1805&  1164&   207 \\\hline
140 -- 160&   1650&  1481&   865&   509&    50 \\\hline
160 -- 200&   1493&  1137&   708&   396&    40 \\\hline
200 -- 240&    503&   406&   242&   107&     6 \\\hline
240 -- 300&    287&   215&   122&    40&     1 \\\hline
300 -- 360&     96&    73&    35&     8&     0 \\\hline \hline
 40 -- 360& 480538&393378&255266&184642& 75660 \\\hline \hline
\end{tabular}
\vskip0.2cm
\caption{Selection 1. $\Delta \Pt^{jet} / \Pt^{jet} \leq 0.10$}
\vskip0.2cm
\begin{tabular}{||c||c|c|c|c|c||} \hline \hline
$\Pt^{\gamma}$  &    HB   &   HB+HE &     HE  &   HE+HF &     HF\\\hline \hline
 40 --  50& 341043& 55160& 263629&26653& 74534\\\hline
 50 --  60& 144955& 20396& 108765& 9300& 29281\\\hline
 60 --  70&  65525&  8541& 49412&  3907& 12621\\\hline
 70 --  80&  34155&  4093& 25918&  1957&  5860\\\hline
 80 --  90&  19224&  1961& 14741&   804&  2778\\\hline
 90 -- 100&  10258&  1304&  8394&   536&  1742\\\hline
100 -- 120&  10859&  1043&  8357&   545&  1338\\\hline
120 -- 140&   4618&   509&  3675&   178&   546\\\hline
140 -- 160&   2325&   222&  1751&    90&   168\\\hline
160 -- 200&   1971&   147&  1458&    52&   147\\\hline
200 -- 240&    685&    61&   472&    20&    26\\\hline
240 -- 300&    383&    32&   234&     7&     9\\\hline
300 -- 360&    129&    10&    72&     1&     0\\\hline \hline
 40 -- 360& 636418& 93480&486788& 44052&129050\\\hline \hline
\end{tabular}
\end{center}
\end{table}

\begin{table}
\vspace{-2mm}
\begin{eqnarray}
\Pt^{isol}_{CUT}=5\;GeV/c; \quad
{\epsilon}^{\gamma}_{CUT}=7\%; \quad
\Delta \phi<15^{\circ}; \quad
\Pt^{clust}_{CUT}=5\;GeV/c.
\end{eqnarray}
No any restrictions on other parameters are used.
The first columns of these tables give the
number of events with jets (found by LUCELL jetfinding algorithm of PYTHIA),
all particles of which
are comprised completely (100$\%$) in the barrel part (HB) and there is a
$0\%$ sharing of $\Pt^{jet}$ ($\Delta \Pt^{jet}$) between HB and neighbouring HE 
part of HCAL, i.e. $\Delta \Pt^{jet}=0$.
Second columns of these tables contain a number of the events in which $\Pt$ of the
jet is shared between HB and HE regions. The same sequence of restriction conditions
takes place in the following columns.
Thus, HE and HF columns include a number of events with the jets 
\begin{center}
\caption{Selection 2. $\Delta \Pt^{jet}/\Pt^{jet}= 0.00$}
\vskip0.2cm
\begin{tabular}{||c||c|c|c|c|c||} \hline \hline
$\Pt^{\gamma}$  &    HB   &   HB+HE &     HE  &   HE+HF &     HF\\\hline \hline
 40 --  50&  46972& 32954& 26114& 16208& 10041\\\hline
 50 --  60&  23717& 18911& 13448&  8367&  5047\\\hline
 60 --  70&  14384&  9751&  7469&  4703&  2386\\\hline
 70 --  80&   8546&  6733&  4627&  2960&  1206\\\hline
 80 --  90&   5653&  4386&  3107&  1925&   573\\\hline
 90 -- 100&   3326&  3119&  1900&  1377&   390\\\hline
100 -- 120&   4157&  3435&  2271&  1467&   324\\\hline
120 -- 140&   2183&  1786&  1185&   710&   134\\\hline
140 -- 160&   1175&  1005&   635&   362&    31\\\hline
160 -- 200&   1179&   905&   565&   314&    25\\\hline
200 -- 240&    442&   353&   212&    97&     5\\\hline
240 -- 300&    273&   200&   116&    37&     1\\\hline
300 -- 360&     94&    71&    35&     7&     0\\\hline \hline
 40 -- 360& 112111& 83617& 61686& 38535& 20163\\\hline \hline
\end{tabular}
\caption{Selection 2. $\Delta \Pt^{jet}/\Pt^{jet}\leq 0.10$}
\vskip0.2cm
\begin{tabular}{||c||c|c|c|c|c||} \hline \hline
$\Pt^{\gamma}$  &    HB   &   HB+HE &     HE  &   HE+HF  &    HF\\\hline \hline
 40 --  50&  60113&  7986& 45388&  3909& 14894\\\hline
 50 --  60&  31495&  3631& 25134&  1971&  7259\\\hline
 60 --  70&  18326&  2248& 13139&   968&  4011\\\hline
 70 --  80&  11385&  1243&  8741&   573&  2132\\\hline
 80 --  90&   7614&   633&  5957&   292&  1145\\\hline
 90 -- 100&   4544&   536&  3886&   280&   865\\\hline
100 -- 120&   5771&   481&  4434&   278&   689\\\hline
120 -- 140&   2909&   272&  2370&    94&   352\\\hline
140 -- 160&   1648&   138&  1246&    65&   111\\\hline
160 -- 200&   1560&   113&  1162&    38&   115\\\hline
200 -- 240&    600&    53&   416&    17&    23\\\hline
240 -- 300&    362&    30&   220&     6&     8\\\hline
300 -- 360&    126&    10&    71&     1&     0\\\hline \hline
 40 -- 360& 146468& 17374&112177&  8492& 31603\\\hline \hline
\end{tabular}
\end{center}
\end{table}
\def\baselinestretch{1.0}

\noindent
completely contained
in these regions, while column HE+HF gives the number of the events where
the jet covers both the HE and HF regions.
From these tables we can see what number of the events
can, in principle, be suitable for the most precise calibration procedure,
carried out separately for HB, HE and HF parts of HCAL in
different intervals of $\Pt^{jet}$.

Less restrictive conditions, when up to $10\%$ of the jet $\Pt$
are allowed to be shared between
HB, HE and HF parts of HCAL, are given in Tables 9 and 11.
Tables 8 and 9 correspond to the case of Selection 1 (see [1]).
Tables 10 and 11 contain a number of events collected with addition of Selection 2
restriction, i.e.
they include only the events with ``isolated jets'' (whose definition is given
by (26) in the same Section 3.2 of [1]).

From Table 8, that corresponds to the most restrictive selection and
gives the number of events most suitable for HCAL calibration, we see
from the last summarizing lines
that for the interval $40<Pt^{\gamma}<360\; GeV/c$ PYTHIA predicts
around a half of million of events for HB and a quarter of million of events
for HE per one month of continuous data taking at
low LHC luminosity, while for HF it is expected to be less than 80 000
events per month.

One should keep in mind that the last columns in Tables 8--11 can not be taken
as the final result here as we have not defined the meaning of sharing
$\Pt^{jet}$ between the HF regions and the region with $|\eta|>5$, i.e. close to a
``beam-pipe'' region. A more accurate estimation can be done here  by finding events
with jets in a wider region than the HF volume restricted by
$3<|\eta^{HF}|<5$ and
 by calculation of a number of those events whose jets are contained in HF
 completely with all particles belonging to them.

\def\baselinestretch{1.0}

\section{~STUDY OF FEATURES OF \gpj EVENTS IN THE HCAL BARREL REGION.}

In this section we shall study a specific sample of events that may be
most suitable for HB calibration. It is a sample of events
in which jets are completely (100$\%$) contained in HB region, i.e.
having 0$\%$ ~sharing of
$\Pt^{jet}$ (at PYTHIA level of simulation) with HE. Below
we shall call them ''HB-events''. A particular
set of these events for $\Pt^{clust}=5\;GeV/c$ is presented in the first column
(HB) of Table 8.
Here we shall use two 
different jetfinders, namely, LUCELL from PYTHIA
and UA1 
(the last one is taken
\footnote{In order to select events with $\Pt^{clust}$ value starting
from $5~ GeV/c$ we have changed the $\Pt$ precluster threshold in the
UA1 
algorithm from 10 
$GeV/c$, taken as default value,
to 1.5 
$GeV/c$
. We also increase the cone radius in the UA1 algorithm from 0.5 to 0.7.}
 from CMSJET [6])
for an equal foot determination of clusters and jets.
The distributions of $\Pt^{clust}$ for generated events found by these two
different jetfinders in two $\Pt^{\gamma}$ intervals,
$40\!<\!\Pt^{\gamma}\!<50\!~GeV/c$
and $\!$ $300\!<\!\Pt^{\gamma}\!<360\!~GeV/c$, are shown in
Fig.~2 for $\Pt^{clust}_{CUT}=30\;GeV/c$.

\subsection{Influence of the $\Pt^{clust}_{CUT}~$ parameter
on the photon and jet $\Pt$ balance and initial state radiation
suppression.}

Here we shall study correlation of $\Pt^{clust}$ with $\Pt^{ISR}$
proportional to $\Pt56$ for the fixed $\langle k_T \rangle$
(The influence of the $\la k_T \ra$ variation on the
\ptgj~ balance is discussed in [4]).

The banks of 1-jet \gpj events gained from the already
mentioned (in Section 2.1) results of PYTHIA
 generation of $10^7$  signal \gpj events  in each of four $\Pt^{\gamma}$
intervals (40 -- 50, 100 -- 120, 200 -- 240, 300 -- 360 $GeV/c$)
will be used here.
The observables defined in  Sections 3.1 and 3.2 of paper [1] will be
restricted here by Selection 1 cuts (17) -- (24)
 of Section 3.2 of [1] and the cut parameters defined by (1).
The luminosity was chosen to be $L=10^{33}\,cm^2sec^{-1}$.

We have chosen two of these intervals  with the
extreme values of $\Pt^{\gamma}$ to illustrate the influence of the
$\Pt^{clust}_{CUT}$ parameter on the distributions of physical variables.
The results of \\[-5.mm]
\begin{floatingfigure}[r]{8cm}
  \hspace{20.0cm} \includegraphics[width=7.8cm]{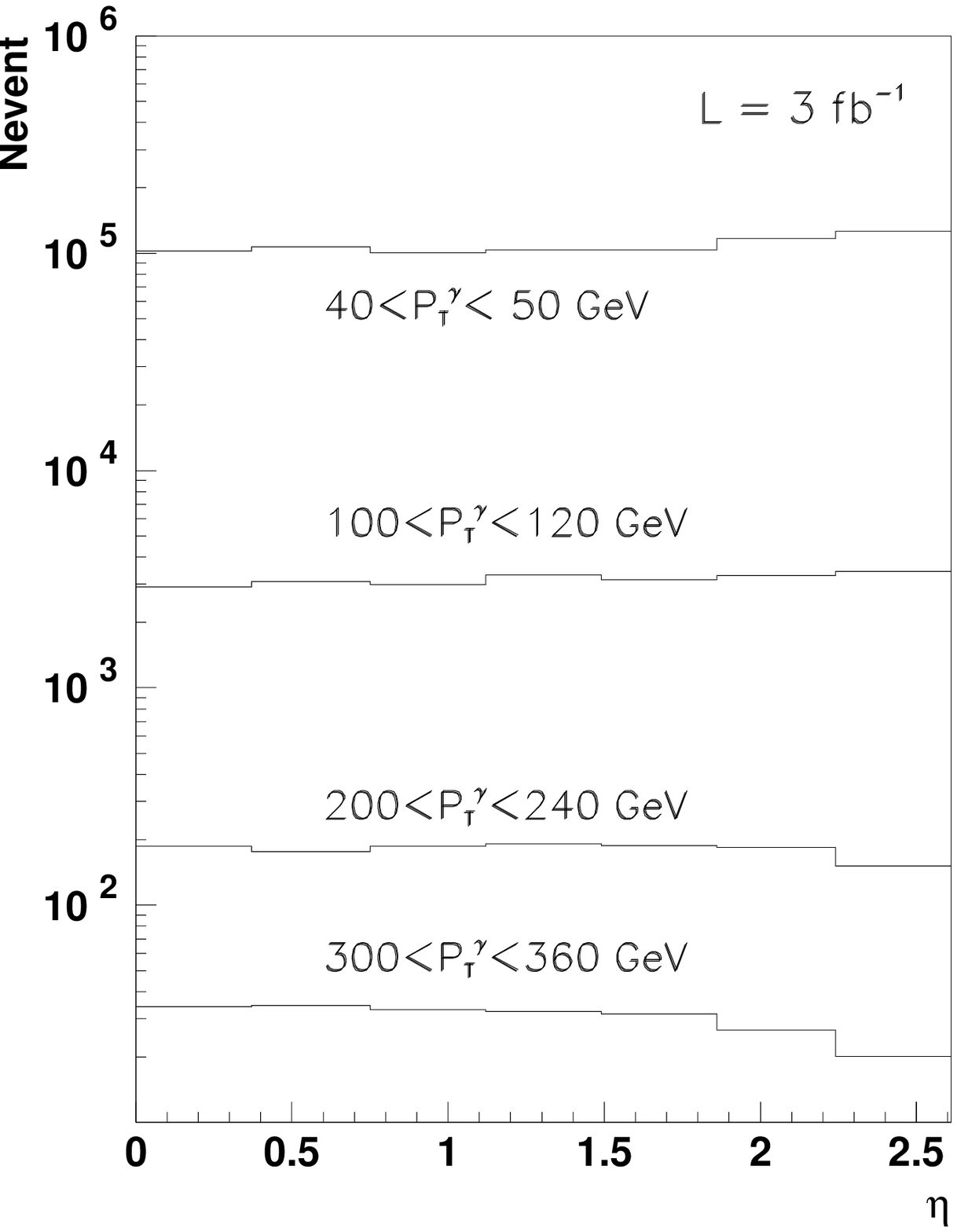}
   \caption{AAA}
    \label{fig7}
  \end{floatingfigure}
\begin{floatingfigure}[r]{8cm}
  \hspace{20.0cm} \includegraphics[width=7.8cm]{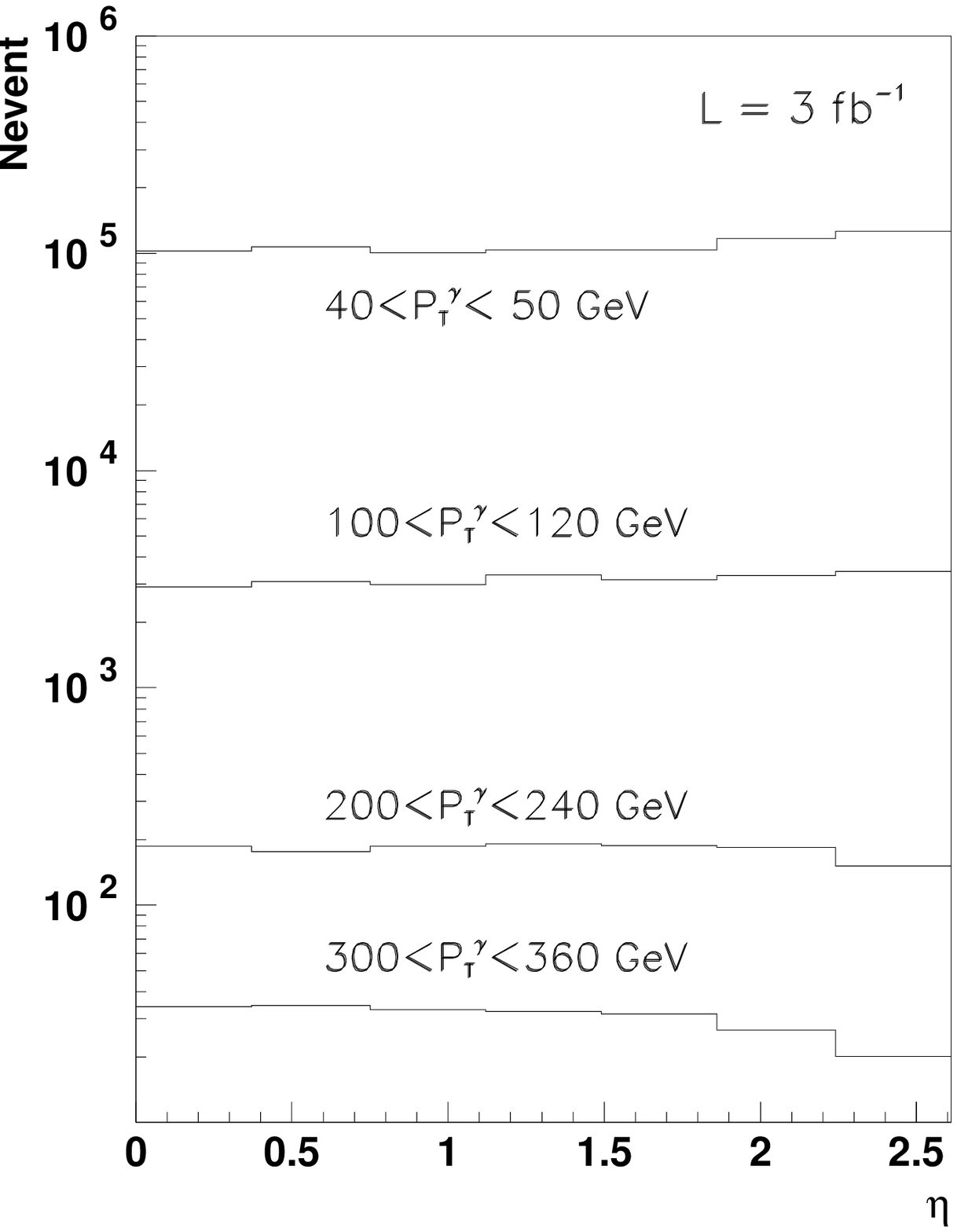}
   \caption{AAA}
    \label{fig8}
  \end{floatingfigure}
\begin{flushleft}
\begin{figure}[h]
 \vspace{-1.1cm}
 \hspace{-.5cm} \includegraphics[height=12.1cm,width=13cm]{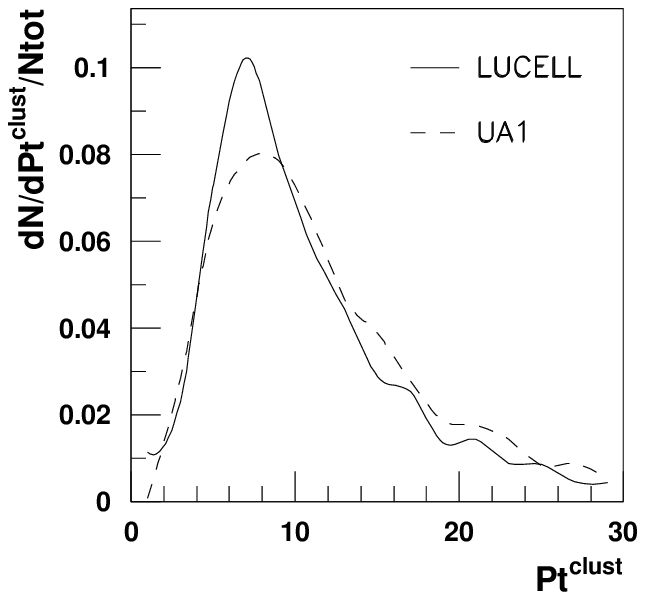}
    \label{fig9}
   \nonumber
  \end{figure}
\end{flushleft}
\begin{flushright}
\begin{figure}[htbp]
 \vspace{-13.9cm}
  \hspace{6.0cm} \includegraphics[height=12.2cm,width=13.3cm]{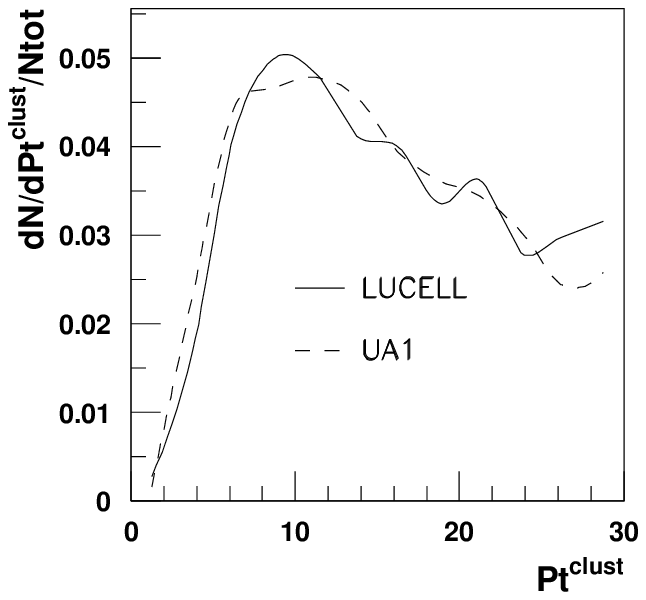}
   \nonumber
\label{fig7}
\end{figure}
\end{flushright}

\vspace*{-7.9cm}
\hspace*{2.0cm} a) \hspace*{6.0cm} b)\\[10pt]
\hspace*{0.5cm}{\footnotesize Fig.~2: $\Pt^{clust}$-distribution in \gpj 
events from two $\Pt^{\gamma}$ intervals:
a) $40<\Pt^{\gamma}<50\, GeV/c$ ~and \\
\hspace*{1.3cm} b) $300<\Pt^{\gamma}<360\,GeV/c$~ with the same cut
$\Pt^{clust}_{CUT}=30\;GeV/c$.}\\

\noindent
$\Pt^{clust}_{CUT}$  variation are shown
in Fig.~3~ for $40<\Pt^{\gamma}<50~ GeV/c$ and in Fig.~4 for
$300<\Pt^{\gamma}<360~ GeV/c$. In these figures, in addition to
three variables $\Pt56$,
$\Pt^{\eta>5}$, $\Pt^{out}$, already explained in Sections 2.2, 3.1 and 3.2
of [1],  we present the distributions for
other two variables, \Db~ and $(1-cos\dphi)$, which
define the right-hand side of equation (29) of [1].
The distribution for
the back-to-back $\dphi$ angle (see (23) of [1]) which defines the second
variable, is also presented in Figs.~3, 4.

The disbalance variable $\Pt56$ (defined by formula (3) of [1])
and both components of another disbalance measure $\Fptgj$
(defined by formula (29) of [1]) as a sum of $(1-cos\dphi)$ and \Db~,
as well as two others, $\Pt^{out}$ and  $\dphi$,
show a tendency, as it is seen from Figs.~3 and 4, to become smaller
 with the restriction of the upper limit on the $\Pt^{clust}$ value.
It means that the calibration precision may increase with decreasing
$\Pt^{clust}_{CUT}$, which justifies the intuitive choice of this new variable in Section 3 of [1]. The origin
of this improvement becomes clear from the
$\Pt{56}$ density plot, which demonstrates suppression of $\Pt{56}$
(or $\Pt^{ISR}$) with implying a more restrictive cut on $\Pt^{clust}$.

The comparison of Fig.~3 (for $~40\!<\Pt^{\gamma}\!<50 ~GeV/c$) and Fig.~4
(for $~300\!<\Pt^{\gamma}\!<360 ~GeV/c$) shows that a degree of
back-to-backness of the photon and jet $\Pt$ vectors in the $\phi$-plane
increases with increasing $\Pt^{\gamma}$. At the same time
$\Pt^{out}$ and $\Pt^{ISR}$ distributions become wider, while
the $\Pt^{\eta>5}$ distribution practically does not depend on
$\Pt^{\gamma}$ and $\Pt^{clust}$.

It should be mentioned that the results presented in Figs.~3 and 4 were
 obtained with the LUCELL jetfinder of PYTHIA
\footnote{The results obtained with both jetfinders and
\ptgj ~balance will be discussed in [2, 3] in more detail}.
~
\begin{center}
\begin{figure}[htbp]
\vspace{-3.0cm}
  \hspace{.0cm} \includegraphics[width=13cm]{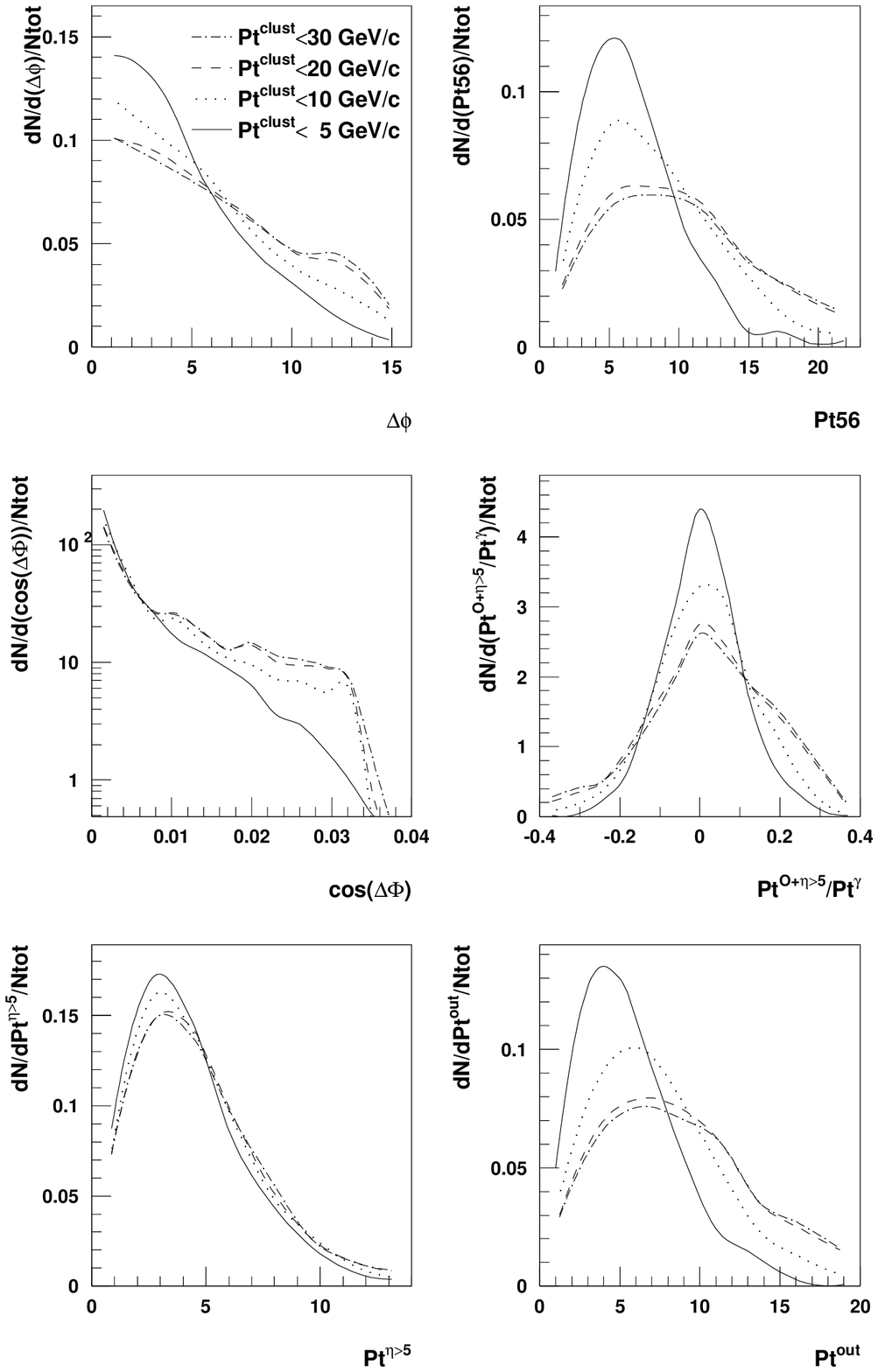}
  \vspace{-0.5cm}
    \caption{\hspace*{0.0cm} LUCELL algorithm, $\dphi<15^\circ$,
$40<\Pt^{\gamma}<50\, GeV/c$. Selection 1.}
    \label{fig8}
\end{figure} 
\begin{figure}[htbp]
 \vspace{-3.0cm}
  \hspace{.0cm} \includegraphics[width=13cm]{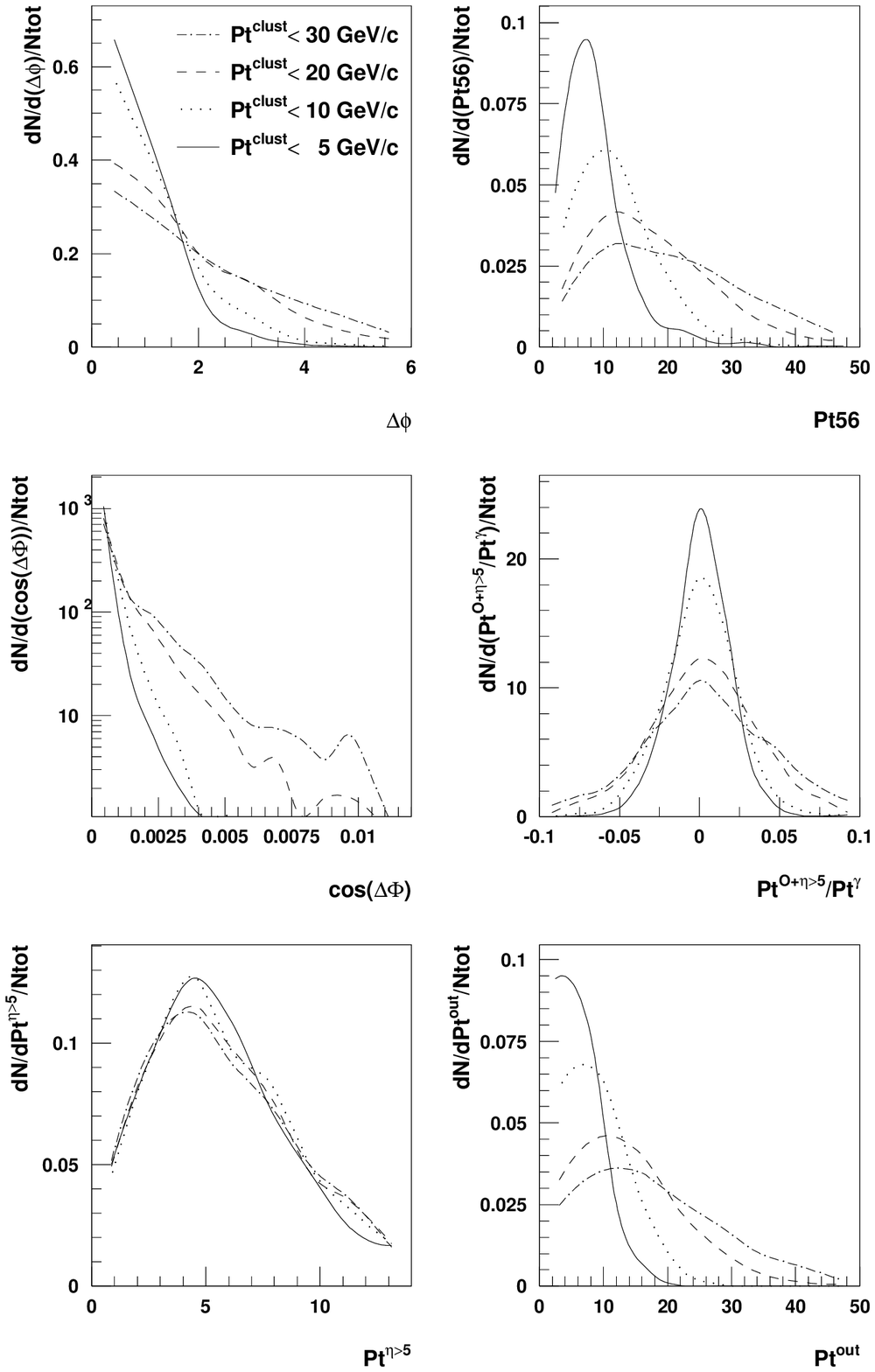}
  \vspace{-0.5cm}
    \caption{\hspace*{0.0cm} LUCELL algorithm, $\dphi<15^\circ$,
$300<\Pt^{\gamma}<360\, GeV/c$. Selection 1.}
    \label{fig12}
\end{figure} 
\end{center}
\subsection{Jetfinders and $\Pt$ structure of jets in $\eta-\phi$ space.}

In order to understand well the calibration procedure of \gpj events,
it is useful to keep control over some principal characteristics of jets
and of the $\Pt$ activity in the space
around them. The left-hand columns of Figs.~5 and 6 represent jet
radius $R^{jet}({\eta,\phi})$ distributions in these HB-events
for the intervals: $40<\Pt^{\gamma}<50\,GeV/c$
and $300<\Pt^{\gamma}<360\,GeV/c$, respectively.
We have chosen the 
jet radius counted from the initiator cell (ic) to be restricted by 
$R^{jet}_{ic}=0.7$ for the LUCELL and UA1 jetfinders.

From the left-hand side plots in Figs.~5 and 6 we see
that both jetfinders give close enough results for $R^{jet}_{gc}$ distribution.
In these plots the radius in $\eta-\phi$ space is counted from the gravity center
(gc). 
The detailed information about
averaged values of the jet radius
for four $\Pt^{\gamma}$-intervals will be presented in the tables of
 Appendices 1--4 of [3]
\footnote{From there one can see a weak dependence of the jet radius on 
$\Pt^{Jet}(\approx\Pt^{\gamma})$
for both UA1 and LUCELL algorithms}.

Let us now consider the question, how the transverse momentum is
distributed inside a jet.
Let us divide the jet radius  $R^{jet}({\eta,\phi})\equiv R$ into a set of
$\Delta R$ bins
and calculate the vector sums of cells $\Pt$ in each $\Delta R_{bin}$ ring.
Normalized to $\Pt^{Jet}$, the modulus of this vector sum, denoted by
$\Pt^{bin}$, would give the value that tells us what portion of
$\Pt^{Jet}$ is contained in the ring of size $\Delta R_{bin}$.
Its variation with the distance $R$ counted from the center of
gravity of the jet is shown in the right-hand columns of Figs.~5 and 6.

From these figures we can conclude that the LUCELL and UA1 jetfinders
lead to equally $\Pt$-densed central part inside the jet. 


\subsection{$\Pt$ distribution inside and outside a jet.}

Now  let us see how the volume outside the jet,
(i.e. calorimeter cells outside the jet cone), may be populated by $\Pt$
in these HB \gpj events. For this
purpose we calculate a vector sum $\vec{\Pt}^{sum}$ of individual transverse
momenta of
$\Delta \eta \times \Delta \phi$ cells, included
by a jetfinder into a jet as well as of cells in a larger
volume that surrounds a jet. In the last case this procedure in some sense
can be viewed  as
some straightforward enlarging of the jet radius in the $\eta -\phi$ space.

The figures that represent the ratio  $\Pt^{sum}/{\Pt^{\gamma}}$,
as a function of the distance $R( \eta,\phi)$ counted from the jet
gravity center towards its boundary and further into space outside the jet
are shown in the left-hand columns of
Figs.~7 and 8 
for a case when all jet particles are kept
in the jet, while the case of the magnetic field effect account 
(i.e. when the contribution
from charged particles with $\Pt \leq 0.8\;GeV/c$ is removed
\footnote{See [5]}
from the total jet $\Pt$ in a case of HB events) 
is shown in the left-hand columns of Figs.~9 and 10.

From these figures we see that the jet surrounding
space is found to be far from being an empty one in the case of
\gpj events considered here. We also see that an average value of
the total $\Pt^{sum}$ increases with increasing the volume around the jet
and it exceeds $\Pt^{\gamma}$
at $R=0.7-0.8$ when all particles are included into the jet
(see Figs.~7 and 8), while in the case of
rejected charged particles
with $\Pt^{ch}<0.8\;GeV/c$ only about
$95-97\%$ of $\Pt^{Jet}$ are collected at $R=0.7-0.8$.

From the right-hand column of Fig.~7 we see that when all
particles are included into the jet, the following disbalance measure:
\begin{center}
\begin{figure}[htbp]
 \vspace{-3.0cm}
  \hspace{.0cm} \includegraphics[width=13cm]{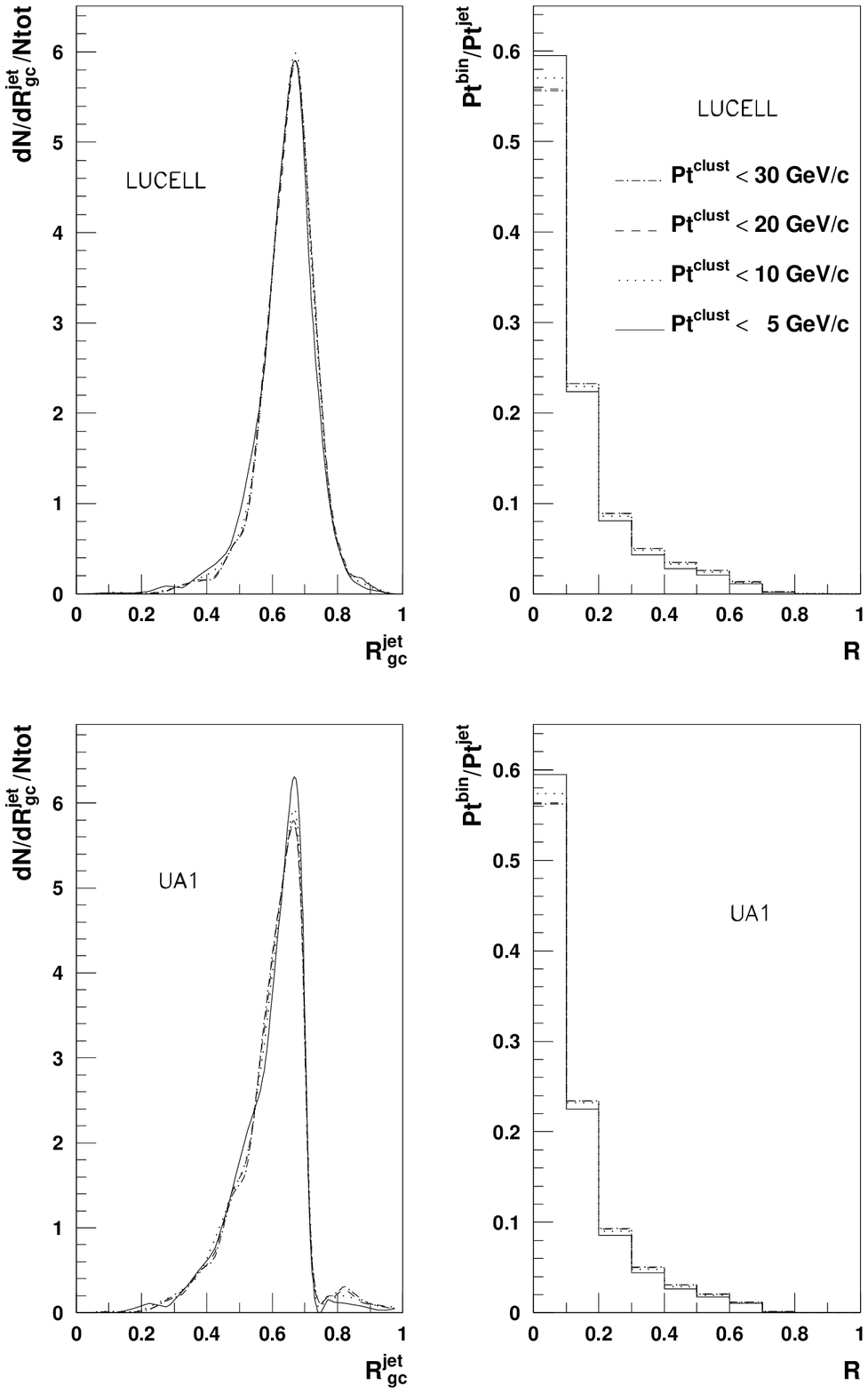}
  \vspace{-0.5cm}
\caption{
\hspace*{0.0cm} LUCELL and UA1 algorithms, $\dphi<15^\circ$,
$40<\Pt^{\gamma}<50\, GeV/c$. Selection 1.}
    \label{fig10} 
  \end{figure}
\begin{figure}[htbp]
 \vspace{-3.0cm}
  \hspace{.0cm} \includegraphics[width=13cm]{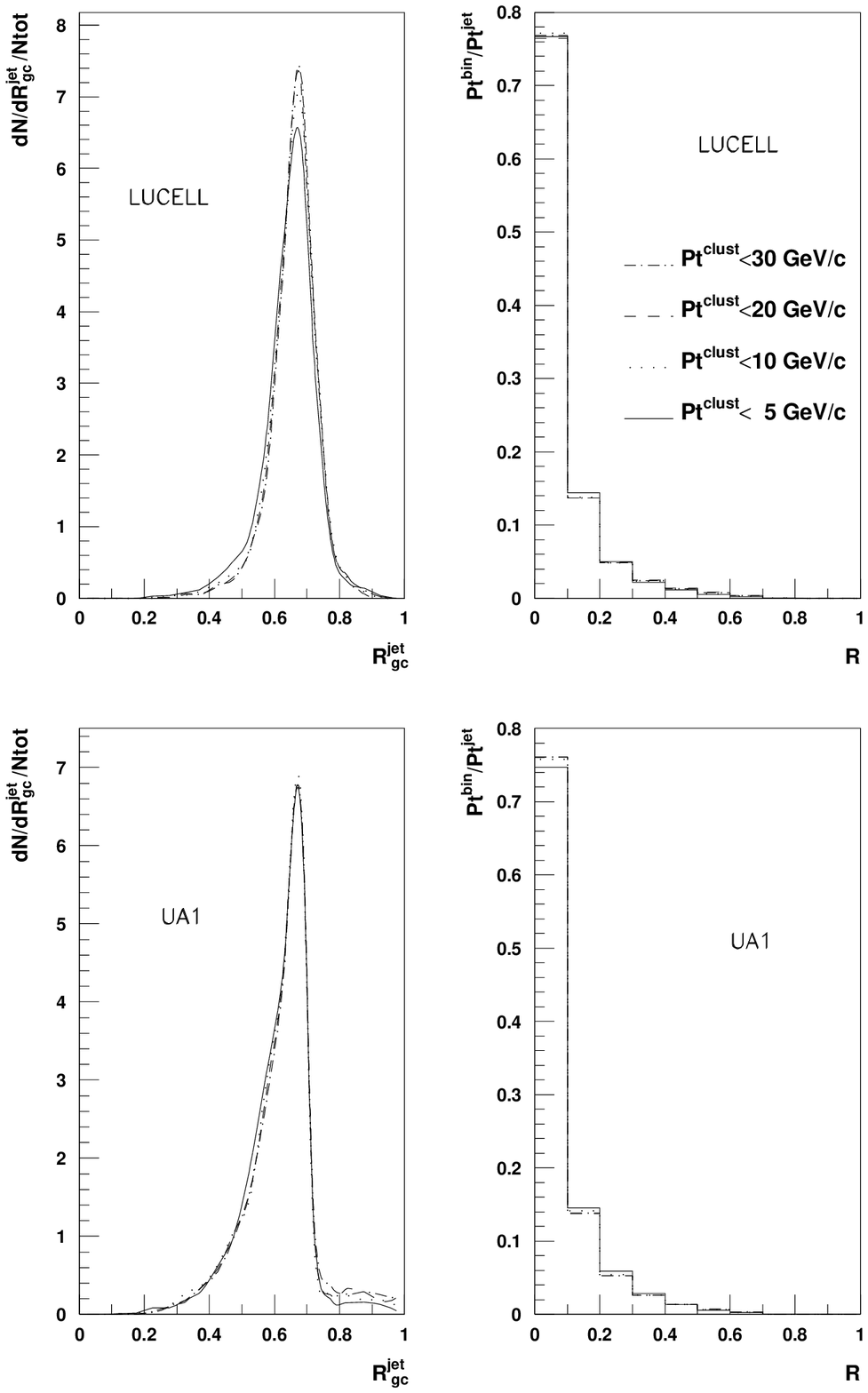}
  \vspace{-0.5cm}
    \caption{
\hspace*{0.0cm} LUCELL and UA1 algorithms, $\dphi<15^\circ$,
$300<\Pt^{\gamma}<360\, GeV/c$. Selection 1.}
    \label{fig11}
  \end{figure}
\begin{figure}[htbp]
 \vspace{-3.0cm}
  \hspace{.0cm} \includegraphics[width=13cm]{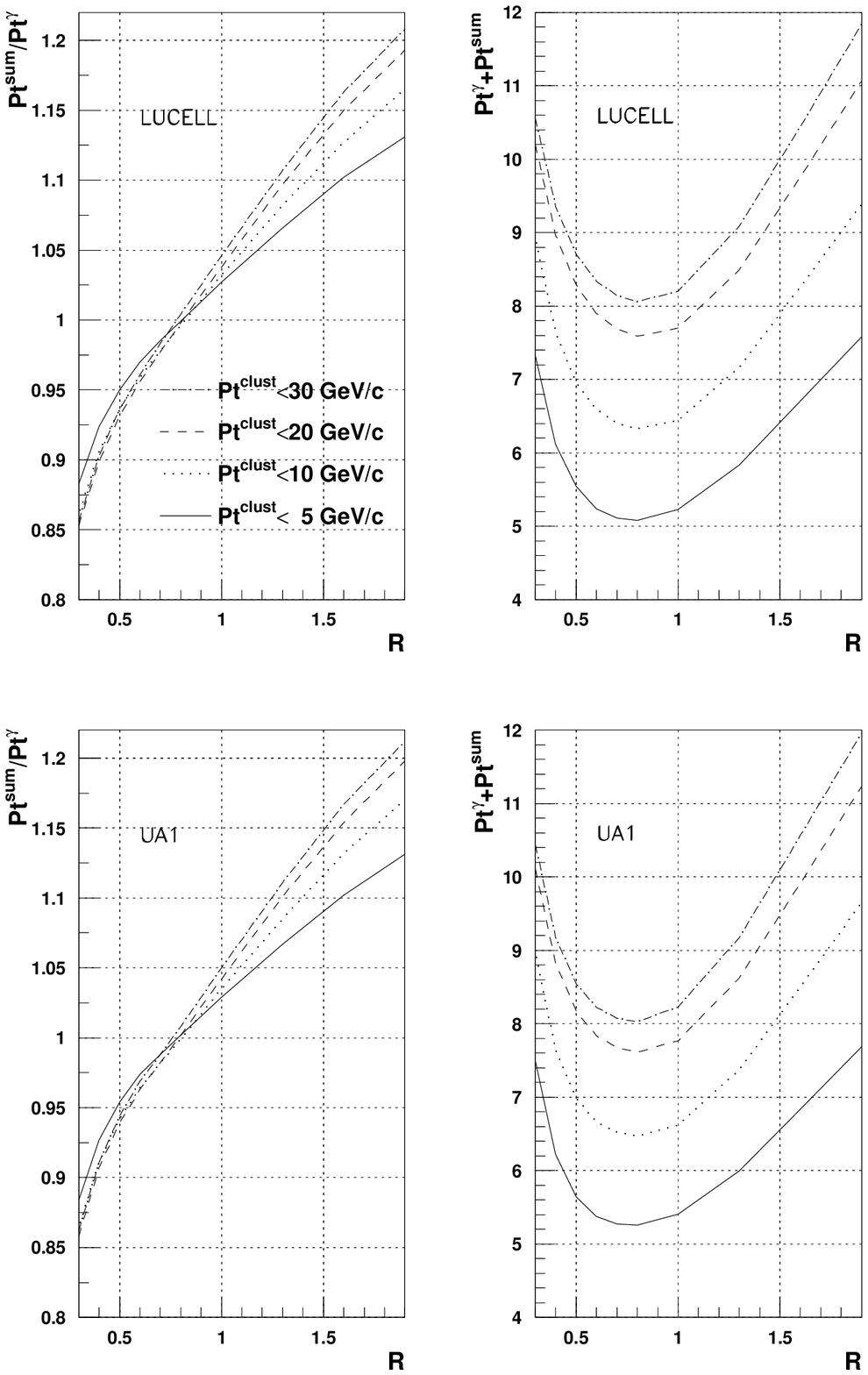}
  \vspace{-0.5cm}
    \caption{\hspace*{0.0cm} LUCELL and UA1 algorithms, $\dphi<15^\circ$,
$40<\Pt^{\gamma}<50\, GeV/c$}
    \label{fig18}
\end{figure}
\begin{figure}[htbp]
 \vspace{-3.0cm}
  \hspace{.0cm} \includegraphics[width=20cm]{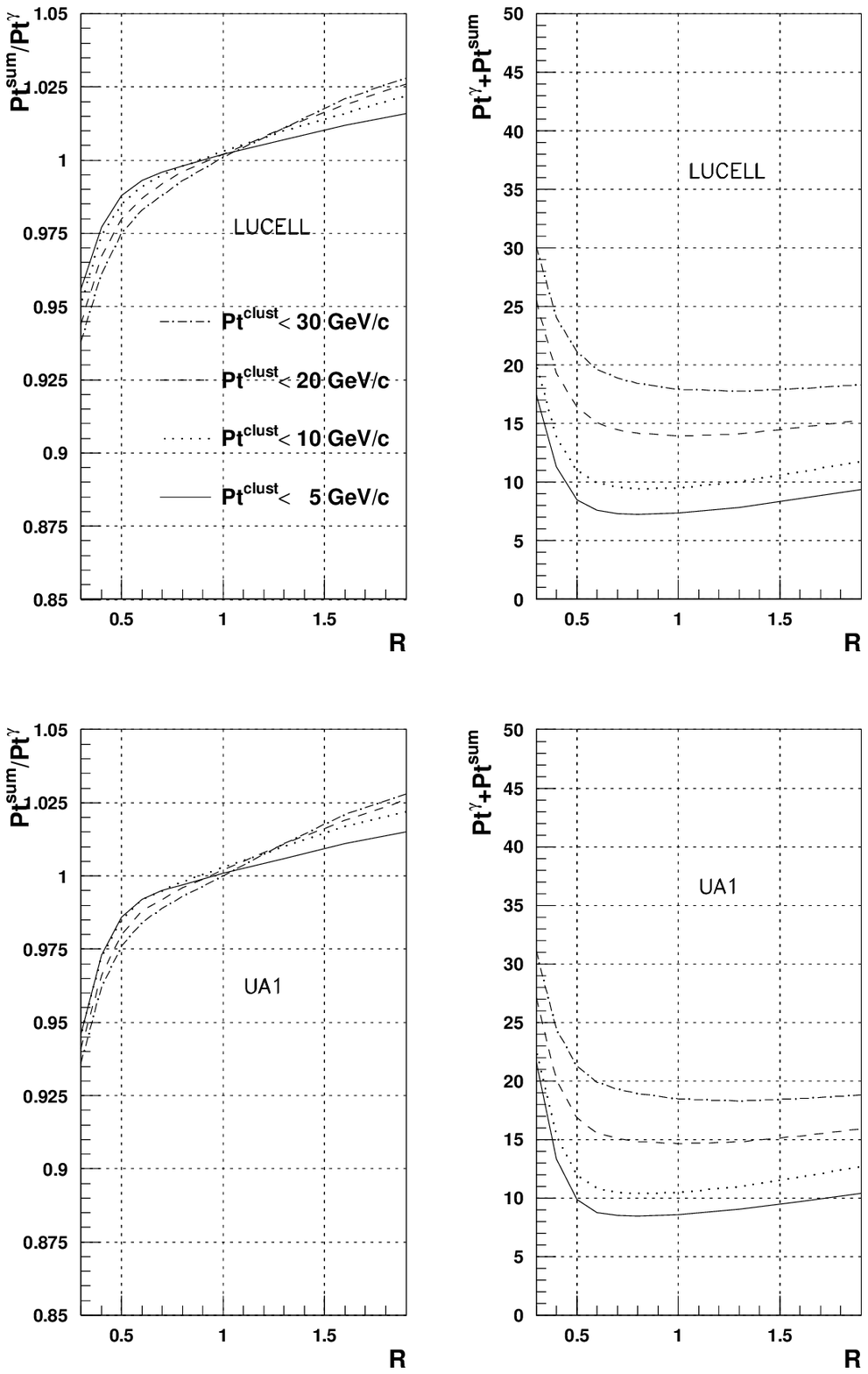}
  \vspace{-0.5cm}
    \caption{\hspace*{0.0cm} LUCELL and UA1 algorithms, $\dphi<15^\circ$,
$300<\Pt^{\gamma}<360\, GeV/c$}
    \label{fig19}
\end{figure} 
\begin{figure}[htbp]
 \vspace{-3.0cm}
\hspace{.0cm} \includegraphics[width=20cm]{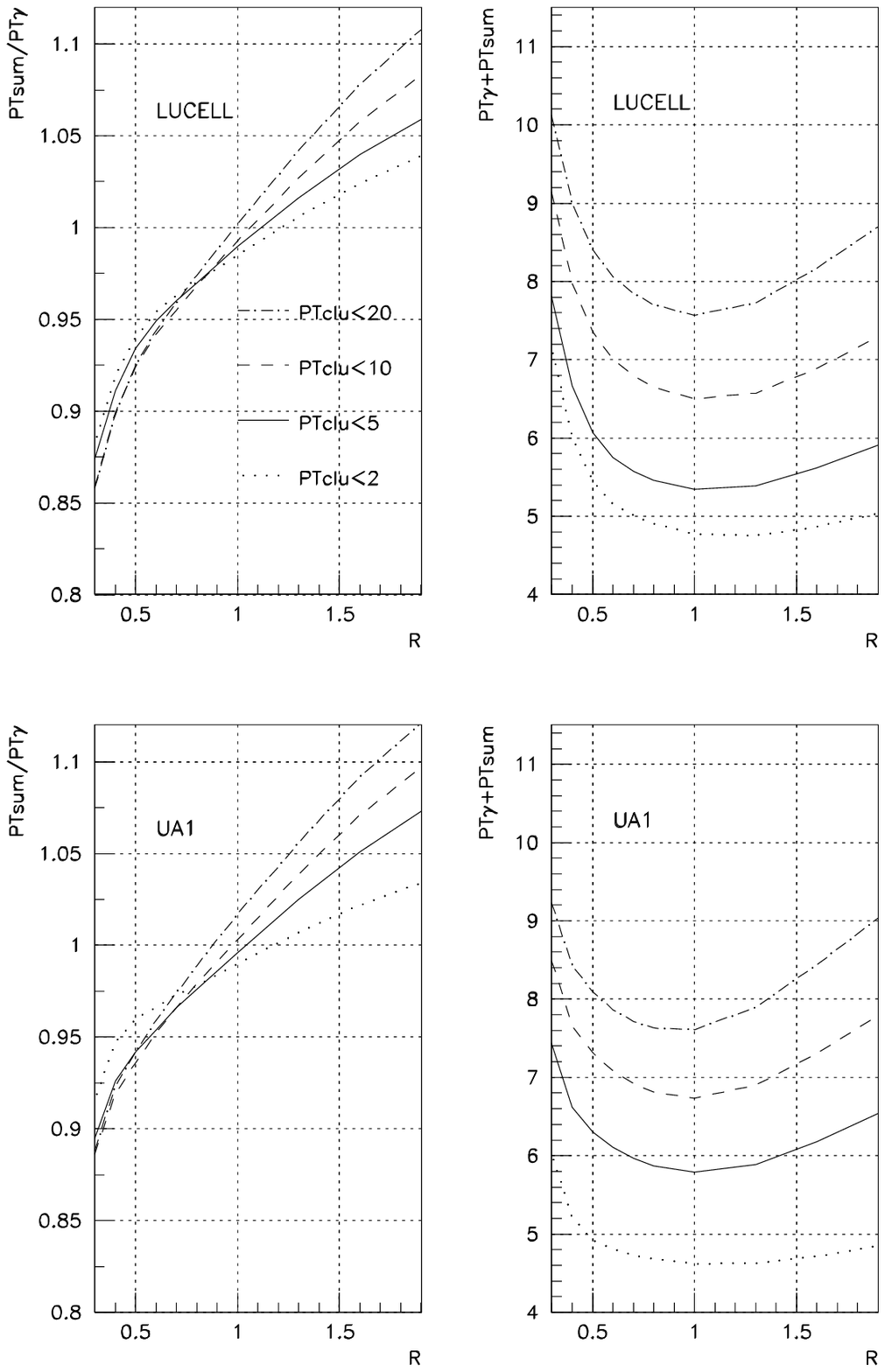}
  \vspace{-0.5cm}
    \caption{\hspace*{0.0cm} LUCELL and UA1 algorithms, $\dphi<15^\circ$,
~$40<\Pt^{\gamma}<50~ GeV/c$,~ $\Pt_{ch}^{jet}>0.8~ GeV/c$.}
    \label{fig16}
\end{figure} 
\begin{figure}[htbp]
 \vspace{-3.0cm}
\hspace{.0cm} \includegraphics[width=13cm,angle=0]{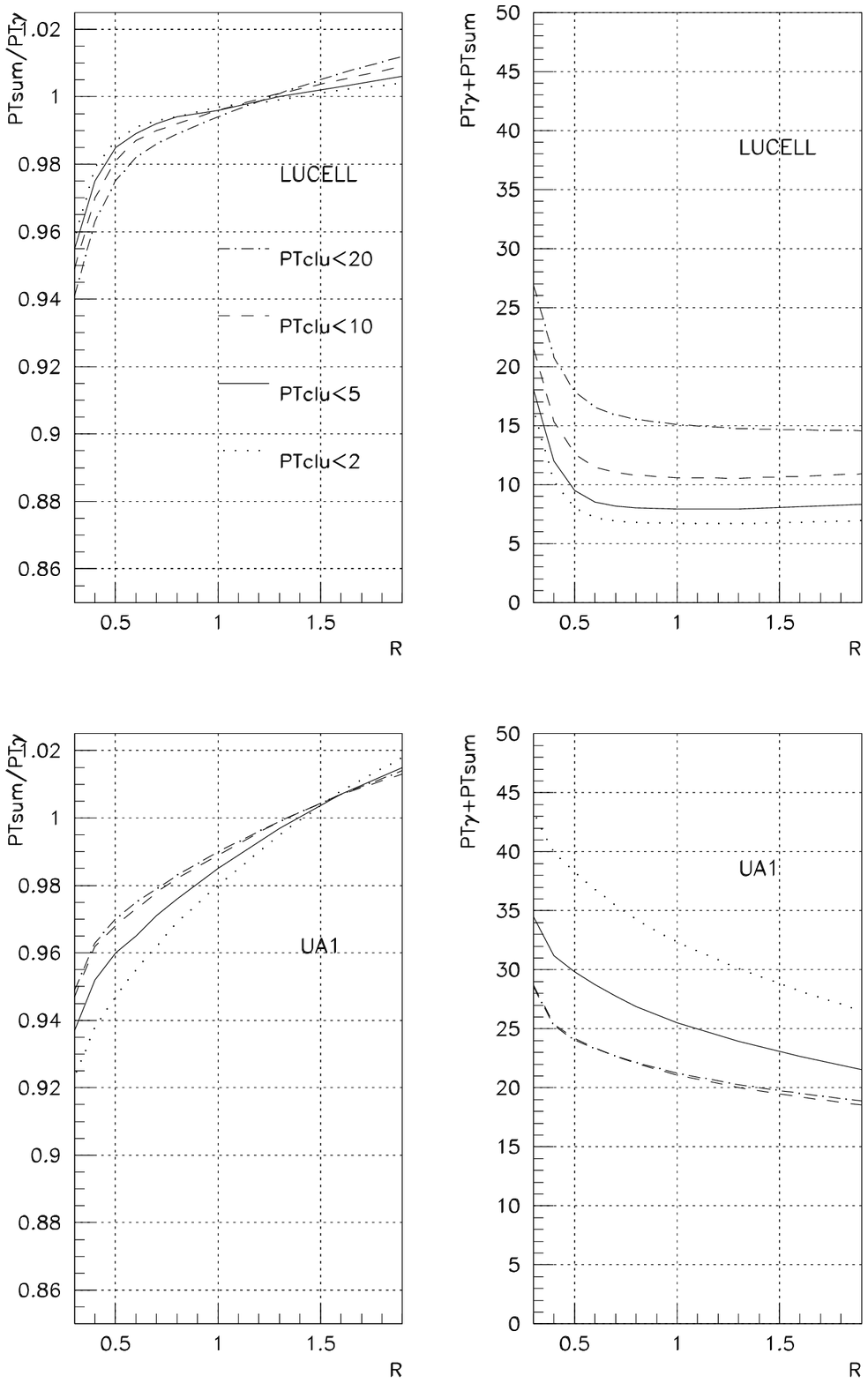}
  \vspace{-0.5cm}
    \caption{\hspace*{0.0cm} LUCELL and UA1 algorithms, $\dphi<15^\circ$,
~$300<\Pt^{\gamma}<360 ~GeV/c$,~ $\Pt_{ch}^{jet}>0.8~ GeV/c$.}
    \label{fig17}
\end{figure} 
\end{center}
~\\[-14mm]
\begin{equation}
\Pt^{\gamma+sum}=
\left|\vec{\Pt}^{\gamma}+\vec{\Pt}^{sum}\right|
\end{equation}
achieves its minimum  at $R\approx 0.7-0.8$ for both LUCELL and UA1
algorithms.
From Figs.~9 and 10 we see that with account of the magnetic
field effect on charged particles with $\Pt^{ch} < 0.8\;GeV/c$~ the minimum of
$\Pt^{\gamma+sum}$ is achieved at larger values of
$R\approx 1.0$ again for both  LUCELL and UA1 jetfinders.

The value of $\Pt^{\gamma+sum}$ continues to grow rapidly with increasing
$R$ after the point $R$=1 for $40<\Pt^{\gamma}<50~ GeV/c$
(see Figs.~7, 9), while for higher
$\Pt^{\gamma}$ (see Figs.~8, 10 for the
$300<\Pt^{\gamma}<360 ~GeV/c$ interval) the ratio
$\Pt^{sum} / \Pt^{\gamma}$ and the disbalance
measure $\Pt^{\gamma+sum}$ increase more slowly with increasing
$R$ after the point $R$=0.7. This means that at higher $\Pt^{\gamma}$
(or $\Pt^{Jet}$) the topology of \gpj events becomes more
pronounced and we get a cleaner picture of an "isolated" jet. This feature
clarifies the motivation of introduction
 of two criteria ``Selection 2'', (see ''Point 8'') and of ``Selection 3''
 (see ''Point 9'') in Section 3.2 of [1]
for selection of events with ``isolated jets''.

\section{SUMMARY}

From the study in Section 2.1 we see that the  PYTHIA simulation predicts
that most \gpj events at LHC energies may have
the vectors $\vec{\Pt}^{\gamma}$ and $\vec{\Pt}^{Jet}$ being back-to-back
within $\Delta\phi<15^\circ$ (about $66\%$ for $40\leq \Pt^{\gamma}\leq 50
~ GeV/c$ and greater than $99\%$ for $\Pt^{\gamma} \geq 200~ GeV/c$).
At the same time, as it is seen from Table 1,
a substantial part of the $\Pt^{ISR}(\approx \Pt56)$ spectrum spreads out in the
interval $0<\Pt56<20 ~GeV/c$, having the peak inside the $5<\Pt56<10 ~GeV/c$
interval and noticeable tails (about of $5-10\%$ of number of events)
extending to $\Pt56=40 ~GeV/c$ not only for $\dphi<15^\circ$ but also
for the smaller $\dphi<10^\circ$ and $\dphi<5^\circ$ intervals.
From here we can conclude that the use of only one traditional
 $\dphi$ cut does not help much to reduce
the $\Pt^{ISR}$ contribution at LHC energies. It is most obvious from
Table 2, where we find that at high $\Ptg$ ($200<\Ptg<240~GeV/c$)
more that $99\%$ of events belong to the $\dphi<15^\circ$ interval.
Most of these events lay within the $0<\Pt56<50 ~GeV/c$ interval.
Even for $\dphi<5^\circ$, a considerable part of the spectrum belongs to
the interval $0<\Pt56<40 ~GeV/c$. Thus, a decrease in $\dphi$ has, in fact,
no big influence on the $\Pt^{ISR}$ spectrum and it cannot be useful
to reduce the number of events with essential amount of
initial state radiation at LHC energies.

In Section 3 the efficiency of $\Pt^{clust}_{CUT}$ for ISR suppression
was demonstrated. From Figs.~3 and 4 we see that
the distribution for
$\Pt56$ becomes narrower with decreasing $\Pt^{clust}_{CUT}$. Analogous
behavior is shown by the $\Pt^{out}$ (as $\Pt^{clust}$ is a part of $\Pt^{out}$)
spectra. These figures serve
as an illustration for a more detailed study of $\Pt^{clust}_{CUT}$ influence
on \ptgj balance, which will be presented in our next papers [2--4].

A strict limitation of the $\Pt^{clust}_{CUT}$ parameter, introduced
in Section 3.2 of [1], (by $\approx 5-10~ GeV/c$)
improves this tendency. Simultaneously, due to this limitation, one can
noticeably reduce radiation in the initial state (compare Tables 1, 4 and
2, 5), which leads to decreasing \ptgj disbalance.

From Section 2.3 (Table 8) we see that even with the restrictive cuts
mentioned there one can expect around a half  million of events for HB,
a quarter million of events for HE, and  less than 80 000 events for
the HF part and $L_{int}=3~fb^{-1}$ (i.e. per month of continuous
data taking at low LHC luminosity)
for interval $40<Pt^{\gamma}<360\; GeV/c$ for the sample of events
that corresponds to $0\%$ $\Pt$ sharing
between HCAL parts. Despite those cuts the number of events for the
HCAL part above, in principle, can be 
quite sufficient for successful jet energy scale setting
and HCAL calibration.\\[-10pt]

\section{ ACKNOWLEDGEMENTS}                                         
We are greatly thankful to D.~Denegri for having offered this theme to study,
fruitful discussions and permanent support and encouragement.
It is a pleasure for us
to express our recognition for helpful discussions to P.~Aurenche,
M.~Dittmar, M.~Fontannaz, J.Ph.~Guillet, M.L.~Mangano, E.~Pilon,
H.~Rohringer, S.~Tapprogge and J.~Womersley.

\end{document}